\documentclass[10pt]{article}
\usepackage{graphicx, verbatim, url, amssymb, amsmath, amsfonts, amsthm, latexsym, hyperref}
\usepackage{caption}
\usepackage{subcaption}

\begin{document}

\title{HITS hits art}

\author{
Massimo Franceschet\\
Department of Mathematics, Computer Science, and Physics \\ 
University of Udine -- Italy \\
\url{massimo.franceschet@uniud.it}
}

\maketitle

\begin{abstract}
The blockchain art market is partitioned around the roles of artists and collectors and highly concentrated among few prominent figures. We hence propose to adapt Kleinberg's authority/hub HITS method to rate artists and collectors in the art context. This seems a reasonable choice since the original method deftly defines its scores in terms of a mutual recursive relationship between authorities/artists - the miners of information/art, and hubs/collectors - the assemblers of such information/art. 

We evaluated the proposed method on the collector-artist network of SuperRare gallery, the major crypto art marketplace. We found that the proposed artist and collector metrics are weakly correlated with other network science metrics like degree and strength. This hints the possibility of coupling different measures in order to profile active users of the gallery and suggests investment strategies with different risk/reward ratios for collectors as well as marketing strategies with different targets for artists. 


\end{abstract}

\section{Introduction} \label{sec:introduction}

Blockchain technology, while commonly associated with cryptocurrencies, has shown potential to bring radical structural change to the arts and creative industries \cite{W19}. Blockchains are already used in the arts including to record provenance and 
authenticity registries \cite{MMPMH17}, to create fractional equity \cite{WK21}, and to guarantee digital scarcity \cite{D18,B17,B18}. 

Another form of blockchain-enabled innovation is \textit{crypto art} \cite{FCSFOSPMH19,F20}. Crypto art is a rising art movement in this cypher space that associates digital artworks with unique and provably rare non-fungible tokens that exist on the blockchain. The real potential of the emerging crypto art current is to give a digital image the dignity of a true work of art, made unique, immutable and collectible through blockchain technology. This created an art market with artists that create and tokenize artworks, collectors that patron or invest on artists, galleries that host marketplaces, as well as curators and art experts that inform the artworks' cultural value through an act of interpretation.

As a significant by-product, crypto art is generating increasing amounts of openly available structured and unstructured data, and this is probably the main feature that sets it apart from traditional art. Indeed, all trades in crypto art are immutably recorded on a public blockchain, and this data is immediately available for analysis. Moreover, artwork metadata like title, description, tags as well as the digital files representing the artworks themselves are stored on peer-to-peer networks like IPFS and available to download. On the contrary, in traditional art this information is typically secreted or available only for a (significant) fee. Besides open data, another facet of crypto art that distinguishes it from its traditional counterpart is velocity. In crypto art something can happen at every instant: an artist mints a new piece or accepts a bid made from a collector, a collector bids or buys and artwork, two users exchange artworks. From a data science viewpoint, the crypto art market corresponds to an open, real-time stream of events, more akin to financial trading than traditional art.

The rating problem in art is to assign an artwork with a given score indicating its extrinsic value (like success of the artwork on the market in terms of bids and sales) as well as its intrinsic value (like the estimation given by an art expert or art curator).
Given a rating score for each artwork in a gallery, we have a ranking of the gallery artworks. Moreover, one can rate and rank artists and collectors by considering them as bags of artworks and extending the score from a single object to a set of objects in some suitable manner. 

The art rating problem is meaningful in a plurality of situations, for instance:

\begin{itemize}
\item a collector wishes to acquire a piece of art for their collection and is looking for a fair estimation of the artwork;
\item a collector wants to insure their art collection or make a will: in both cases they need a rating  of the collection;
\item an investor wants to diversify investments in the field of art and hopes to identify a set of art pieces with potential optimal return on investment;
\item an auction house needs an estimation of a piece of art in view of an incoming auction;
\item an artist longs for an assessment of an artwork they created in order to fix the reserve price for an incoming auction;
\item a notable collector or an important artist yearns for a scrupulous evaluation of their art collection in order to mint a personal fungible token backed by their collection. 
\end{itemize}

Presently, there exists no standard nor any shared proposal for defining ratings for crypto artists and artworks. This despite the public availability of all data for crypto art on blockchains and other public peer-to-peer networks (like IPFS). Is this contribution, we approach the rating problem - at least the valuation of the extrinsic value of an artwork - with network centrality methods. We start from the intuition that important art collectors buy from important artists and important artists sell to important art collectors. We adapt the Hyperlink-Induced Topic Search (HITS) method \cite{K98}, originally developed to detect hubs and 
authorities on the Web, to rate and rank artists and collectors in art systems. We propose to run the  HITS method on a (weighted) collector-artist network that links a collector with an artist if the collector bought an artwork of the artist for a given price. We associate HITS authorities with artists and HTIS hubs with collectors: artists create and sell artworks, they are the sources of 
art. Collectors buy and pull together artworks, they have some sense of where good art is. We apply the proposed rating method to the marketplace of SuperRare, which is among the most important crypto art galleries by popularity and volume of exchanged artworks. 

The rest of the paper is organized as follows. In section \ref{sec:blockchain} we briefly introduce blockchain technology and its application to art. Section \ref{sec:the-rating-problem} sketches the art rating problem while we propose our solution in the next Section \ref{sec:hits}. Section \ref{sec:apply} applies the solution to SuperRare dataset. We draw conclusions in the last Section \ref{sec:conclusion}.

\section{Blockchain art} \label{sec:blockchain}

The origins of the blockchain go back to the \emph{crypto-anarchism} and \emph{cypherpunk} movements of the late 1980s. These activists advocated the widespread use of strong cryptography to guarantee confidentiality and security while sending and receiving information over computer networks, in an effort to protect their privacy, their political and economic freedom \cite{M88}. The following excerpt from the \emph{Cypherpunk Manifesto} by Eric Hughes is particularly telling since it contains, some 30 years in advance, all the ingredients that inspire modern blockchain technology \cite{H93}:

\begin{quote}
\textit{We the Cypherpunks are dedicated to building anonymous systems. We are defending our privacy with cryptography, with anonymous mail forwarding systems, with digital signatures, and with electronic money. [...] Cypherpunks write code. We know that software can't be destroyed and that a widely dispersed system can't be shut down}.
\end{quote}

Blockchains are hard to grasp at first. The basic scientific research from which the technology emerged -- a journal paper and a US patent of Stuart Haber, a cryptographer, and Scott Stornetta, a physicist \cite{HS91,HS92} -- is distinct from the financial systems it later generated -- the advent of bitcoin and other cryptocurrencies \cite{N08,V14}.

Haber and Stornetta were trying to deal with epistemological problems of how we trust what we believe to be true in a digital age \cite{W19,HS91}:

\begin{quote}
\textit{The prospect of a world in which all text, audio, picture and video documents are in digital form on easily modifiable media raises the
issue of how to certify when a document was created or last changed. The
problem is to time-stamp the data, not the medium.}
\end{quote}

In particular, they started from two questions \cite{W19}:

\begin{enumerate}
\item
  \begin{quote}
\textit{If it is so easy to manipulate a digital file on a personal computer,
  how will we know what was true about the past?}
\end{quote}
\item
  \begin{quote}
\textit{How can we trust what we know of the past without having to trust a
  central authority to keep the record?}
\end{quote}
\end{enumerate}

These questions configure an extremely challenging problem. The problem was solved by Haber and Stornetta \cite{HS91} and, 17 years later, by Satoshi Nakamoto\footnote{Satoshi Nakamoto is the pseudonymous used by the person or persons who developed bitcoin, authored the bitcoin white paper, and created and deployed bitcoin's original reference implementation.} \cite{N08} as well, using a combination of tools borrowed from mathematics, computer science, economics and political science.

In their original, far-sighted proposal, Haber and Stornetta  envisaged the adoption of blockchains beyond texts, and maybe in the context of art as well \cite{HS91}:

\begin{quote}
\textit{Of course digital time-stamping is not limited to text. Any string of bits can be time-stamped, including digital audio recordings, photographs, and full-motion videos. {[}\ldots{}{]} time-stamping can help to distinguish an original photograph from a retouched one.}
\end{quote}

Indeed, blockchain technology, while commonly associated with cryptocurrencies, has the potential to bring radical structural change to the arts and creative industries. The blockchain has core use cases in the arts including \emph{provenance}, \emph{fractional ownership} and \emph{digital scarcity}. A notable example of the first use case -- provenance -- is the sale in 2018 of the Barney 
A. Ebsworth collection at Christie's for 318M\$. The auction was held in partnership with the technology provider Artory using a blockchain solution to record information about the auction and all future sales of the auctioned artworks. As for fractional ownership, in 2018 the company Maecenas bought Andy Warhol's 14 Electric Chairs and divided it up into shares sold as so-called ART tokens. The company raised 1.7M\$ for 31.5\% of the artwork at a valuation of 5.6M\$.

Crypto art is related to the third use case of blockchain in art: digital scarcity \cite{FCSFOSPMH19}. The novel idea is to make a digital file scarce by associating it with a \emph{non-fungible token} or NFT \cite{NADMAB21,WLWC21}. A \textit{cryptographic token} is a quantified and tradable unit of value recorded on the blockchain. There are two types of tokens:

\begin{enumerate}
\item \textit{fungible tokens} are cryptocurrencies like Bitcoin and alternative coins. They are interchangeable and can be split in smaller pieces whose sum makes the whole;

\item \textit{non-fungible tokens} (NFTs) represent something unique. You may think of them like rare, one-of-a-kind collectibles. They are not interchangeable and cannot be divided.
\end{enumerate}

Three major applications of NFTs are blockchain gaming, digital land and digital art. For instance, Sandbox\footnote{\url{https://www.sandbox.game}} is a virtual world built on the Ethereum blockchain, where players can build, own, and monetize their gaming experiences. The metaverse has a SAND token, a fungible token that is used for value transfers as well as staking and governance. The Sandbox offers an asset marketplace where virtual assets (published as NFTs) are bought and sold for SAND. As for digital land, Decentraland\footnote{\url{https://decentraland.org}} and Cryptovoxels\footnote{\url{https://www.cryptovoxels.com}} are decentralized virtual reality worlds (metaverses) where players can own and exchange pieces of virtual land and other in-game NFT items. Virtual land is associated with a NFT that can be traded or even put on rent. On a virtual land you can, for instance, open a digital art gallery and display your own collection of digital artworks (which are themselves NFTs).

Crypto art is digital art minted and traded on the blockchain. In crypto art, a NFT certifies the scarcity (number of copies), ownership (current owner) and provenance (historical owners and creator) of a digital artwork. Transferring the NFT is akin to transferring the certificate of ownership of the artwork. However, like in traditional art, ownership rights generally do not include intellectual property rights such as copyright claims and rights for any commercial re-use. Crypto art draws its origins from conceptual art, sharing the immaterial and distributive nature of artworks, and the rejection of conventional art markets and institutions \cite{F18}. A niche artistic movement until early 2020, crypto art market went parabolic in late 2020 - also because of COVID pandemic - attracting the attention of major mass media and major auction houses. Recent notable sales of crypto artworks include:

\begin{enumerate}
\item
\textit{Everydays: The First 5000 Days}, by digital artist Beeple, was the first NFT sold at Christie's on March 2021 for the 
record-breaking amount of 69M\$ through the crypto art gallery MakersPlace;
\item
\textit{The Fungible}, by digital artist Pak, is an NFT collection sold in April 2021 at Sotheby's in collaboration with crypto art gallery Nifty Gateway for almost 17M\$;
\item 
Nine CryptoPunks from Larva Labs' own collection sold in May 2021 at Christie's for 16.9M\$. It is, unsurprisingly, the first time an NFT has been offered alongside work by Andy Warhol and Jean-Michel Basquiat. 
\end{enumerate}

The typical workflow of crypto art can be illustrated with respect to the digital gallery SuperRare\footnote{\url{https://superrare.co}}:

\begin{enumerate}
\item
an artist creates a digital artwork and uploads it to the gallery. The author specifies the title,  description, a list of tag words and possibly a price;
\item
the smart contract of the gallery creates a NFT on the Ethereum blockchain associated with the artwork, and transfers the token to the artist's digital wallet\footnote{Ethereum is a public blockchain featuring a smart-contract (scripting) functionality. A smart contract is a computerized transaction protocol that executes the terms of a contract. Ether (ETH for short) is the cryptocurrency generated by the Ethereum platform as a reward to mining nodes. A digital wallet is a software that allows blockchain users to manage and securely store their own private keys instead of recording them manually.};

\item the gallery distributes the artwork file over the IPFS peer-to-peer network\footnote{The InterPlanetary File System (IPFS, \url{https://ipfs.io}) is a protocol and peer-to-peer network for storing and sharing hypermedia in a distributed file system. IPFS uses content-addressing storage to uniquely identify each file, a way to store information so it can be retrieved based on its content, not its location. Each file is identified by the hash of its content. IPFS lets you address large amounts of data and place permanent links into blockchain transactions.}; hence neither the token nor the artwork are on any central server;

\item
  collectors can place bids on the artwork by transferring the bid
  amount to the smart contract of the gallery (the collector can
  withdraw bids at any time) or they can buy directly at the
  price set by the artist;
\item
  eventually, the artist accepts a bid; the smart contract of
  the gallery then transfers the artwork's token to the collector's wallet
  and the agreed amount of cryptocurrency to the artist's wallet;
\item
the artwork remains on the market. Each re-sale in the secondary market\footnote{The art market is split up into the primary and secondary market. The primary art market is when the price for the piece of art is established for the first time and the artwork is sold from the creator to the first collector. The secondary market usually trades in established and sought-after artists. Once a piece has been acquired on the primary market and is being re-sold, it is now part of the secondary market or second-hand market.} of the SuperRare gallery rewards also the original artist \cite{WK21}.
\end{enumerate}

Finally, it is instructive to cast a parallel between the \emph{traditional} and \emph{crypto} art markets. Crypto art differs markedly from the traditional art market on two  dimensions of crucial importance: the lack of an ecosystem mediating access to opportunities, and the full availability/transparency of market data.  In traditional art markets, most of the market data is not available. This, we argue, is a factor in determining the existence and importance of an ecosystem to broker opportunities, which would be less influential given price transparency. In crypto art, all market data are stored on blockchains, typically Ethereum, and thus are fully available. Nevertheless, different crypto art galleries and marketplaces still use a variety of standards to register their data, therefore inter-gallery price transparency, not to mention interoperability, are still limited.

\section{The rating problem in art markets} \label{sec:the-rating-problem}

In traditional art, the primary and secondary markets have been separated. The primary market, galleries in particular, work with a selection of artists that they promote over time. The main pricing system that galleries use is based on \textit{pricing scripts} \cite{V07}. An artwork is priced using what in practice is a simple linear model mainly including values related to the materials and technique, the size and a multiplier related to the artist's estimated market value \cite{RV02,SR07}. The logic of a gallery is that of curating the price of an artist over time through social and interpersonal relations. In the secondary market, instead, the pricing of artworks happens primarily via auctions, in view of maximizing profits according to supply and demand \cite{AG03,MM05}. 

Methods for art price estimation have mostly been developed for the purpose of constructing price indexes for investment. Two families of methods have emerged from economics: repeat-sales regression and hedonic regression \cite{GMM06}. Repeat-sales regression uses the prices of the same object traded at two or more points in time. Hedonic regression, instead, regresses prices on characteristics of artworks (e.g., size, artist, style, and more) and uses the regression residuals to compute price indexes. While repeat-sales regression allows to bypass the issue of measuring the heterogeneous characteristics of artworks entirely, hedonic regression allows to estimate the price of artworks in the absence of a previous trading history. 

Presently, there exists no standard nor any shared proposal for defining ratings for crypto artists and artworks. This despite the public availability of all data for crypto art on blockchains and other public peer-to-peer networks (like IPFS). Is this contribution, we concern ourselves with the pricing of art on any market, either primary or secondary. In particular, we propose a method to establish a rating for artists coupled with a rating for collectors, calculated independently from the characteristics of artworks, which can be hard to measure, and from re-sale history, which is often absent. Our rating system can therefore be used as the artist multiplier when applying pricing scripts in a gallery setting, and as a model variable for hedonic regression when considering price indexes or auctions.

\section{Rating artists and collections in crypto art} \label{sec:hits}

To define a rating for crypto artists and collectors, we focus on the mutual relationship between artists and collectors. We start from the intuition that important collectors buy from important artists and important artists sell to important collectors. We borrow the Hyperlink-Induced Topic Search (HITS) method \cite{K98}, originally developed to detect hubs and authorities on the Web, and adapt it to the art context. 

HITS assumes that in certain networks there can be found two types of important nodes \cite{K98}: \emph{authorities}, that contain
reliable information on the topic of interest, and \emph{hubs}, that tell us where to find authoritative information. A node may be an authority, a hub, both or neither. For instance, on the Web hubs are pages that compile lists of resources relevant to a given topic of interest, while authorities are pages that contain explicit information on the topic. In an article citation network, hubs are for example review papers that mainly reference other papers containing relevant information on a given topic, while authorities are articles that contain the explicit information. This calls for two distinct yet interrelated notions of centrality: authority and hub centrality. There is a mutual recursion underlying the definition of the roles of authorities and hubs that can be concisely expressed as follows:

\begin{quote}
A node is an authority if it is linked to by hubs (nodes with high hub centrality); a node is a hub if it links to authorities (nodes with high authority centrality).
\end{quote}

Formally, let \(A\) be the adjacency matrix of a directed network. The authority centrality \(x_{i}\) of node \(i\) is proportional to the hub centrality of the nodes that link to it, that is:

\[x_i = \alpha \sum_k A_{k,i} \, y_k\]

On the other hand, the hub centrality \(y_{i}\) of node \(i\) is proportional to the authority centrality of the nodes linked by it, that is:

\[y_i = \beta \sum_k A_{i,k} \, x_k\]

where \(\alpha\) and \(\beta\) are constants. If the network is weighted, then \(A_{i,j}\) is a positive number that represents the
strength of the relationship between nodes \(i\) and \(j\): the higher the weight, the stronger the link. Notice how the above equations use these weights: stronger links give more (authority and hub)
centralities. In matrix form the above equations write:

\[\begin{array}{lcl}
x & = & \alpha y A \\
y & = & \beta x A^T \\
\end{array}\]

In this formulation the mutual reinforcement between hubs and authorities is evident: authorities (\(x\)) depend on hubs (\(y\)) and hubs (\(y\)) depend on authorities (\(x\)), with the mediation of the network structure encoded in matrix \(A\).

In the crypto art context, we associate authorities with artists and hubs with collectors. Artists create and sell artworks, they are the sources of art. Collectors buy and pull together artworks, they have some sense of where good art is. We can hence rephrase Kleinberg's thesis in the art setting as follows:

\begin{quote}
A leading artist sells to leading collectors and a leading collector buys from leading artists.
\end{quote}

Our rating proposal is therefore the following. We start by building a \emph{collector-artist directed network} as follows. The nodes of the network are the active users of a marketplace, that is those users that made at least one sale, one purchase or one mint of a piece of art. The links are drawn as follows. Suppose there is a sale in which a given user sells to a collector C an artwork for a price P. Let A be the artist that originally created the sold artwork (in case of primary sale, the seller is the artist itself). We then add to the network a link from C (the collector) to A (the artist) weighted by the price P. In fact, we weigh the link with the amount paid by the collector in USD, using the ETH-USD exchange rate of the day of the transaction. This link represents a weighted endorsement made by collector C to artist A.

On the collector-artist network we compute the following \textit{node centrality measures}:

\begin{enumerate}
\item
\emph{in-degree}: the unweighted in-degree of a node, which corresponds to the number of artworks created by the node that were sold on either the primary or secondary market of the gallery;
\item
\emph{out-degree}: the unweighted out-degree of a node, which corresponds to the number of artworks bought by the node on either the primary or secondary market of the gallery;
\item
\emph{in-strength}: the weighted in-degree, which corresponds to the overall amount (in USD) made by sales of artworks created by the node on either the primary or secondary market of the gallery;
\item
\emph{out-strength}: the weighted out-degree, which corresponds to the overall amount (in USD) spent by the node on either the primary or secondary market of the gallery;
\item
\emph{authority}: the Kleinberg's HITS authority rating, either on the weighted or the unweighted collector-artist network;
\item 
\emph{hub}: the Kleinberg's HITS hub rating, either on the weighted or the unweighted collector-artist network;
\end{enumerate}

It is worth noticing that a characteristic of the art market, one that allows to draw a parallelism with the scientific publication system, is the \emph{mechanism of endorsement} of artists and collectors. Both works of art and science can be endorsed by the respective communities, thus gaining in popularity and, for artworks, in commercial value. A scientific paper (author) is endorsed when a peer references it in another article. An artwork (artist) is endorsed when a collector buys it. The number of times the artwork is traded among collectors might indicate the \emph{popularity} of the piece of art in the artistic setting, as much as the number of citations from other scholars accrued by a paper is an indicator of its popularity within a scientific community. Furthermore, besides popularity, one can also investigate the \emph{prestige} of the works of art and of scholarly publications and, indirectly, of artists and authors. We might argue that a sale of an artwork to a prestigious collector, or a citation to an article given by an authoritative scientist, are more important than endorsements given by a minor collector or scientist.

\section{Application to SuperRare dataset} \label{sec:apply}

\begin{figure}[t]
	\begin{center}
  \includegraphics[scale = 0.5]{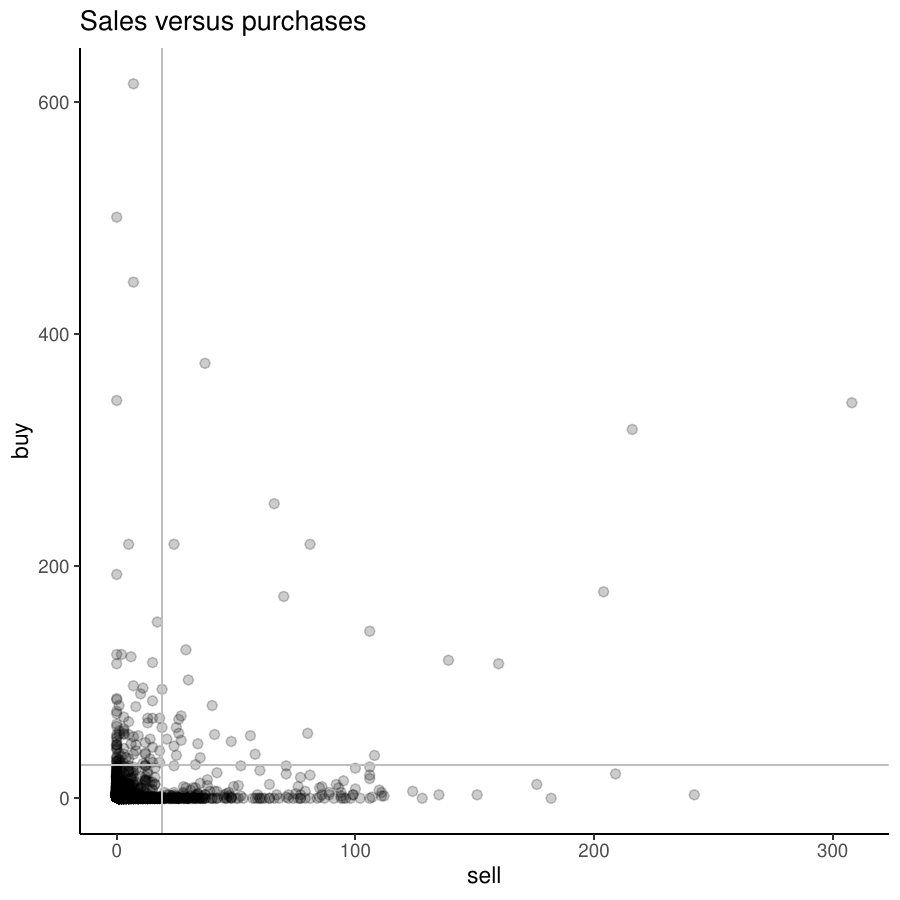}
  \end{center}
	\caption{A scatterplot of number of sales and purchases. Each point is a user of SuperRare and the position of the point is determined by the number of sales and purchases made by the user. The vertical and horizontal lines are the 95th percentile of the sell and buy dimensions. These lines partition the plot in four zones corresponding to roles of users (read more in the text).}
	\label{fig:sell-buy}
\end{figure}

\begin{figure}[t]
	\begin{center}
  \includegraphics[scale = 0.35]{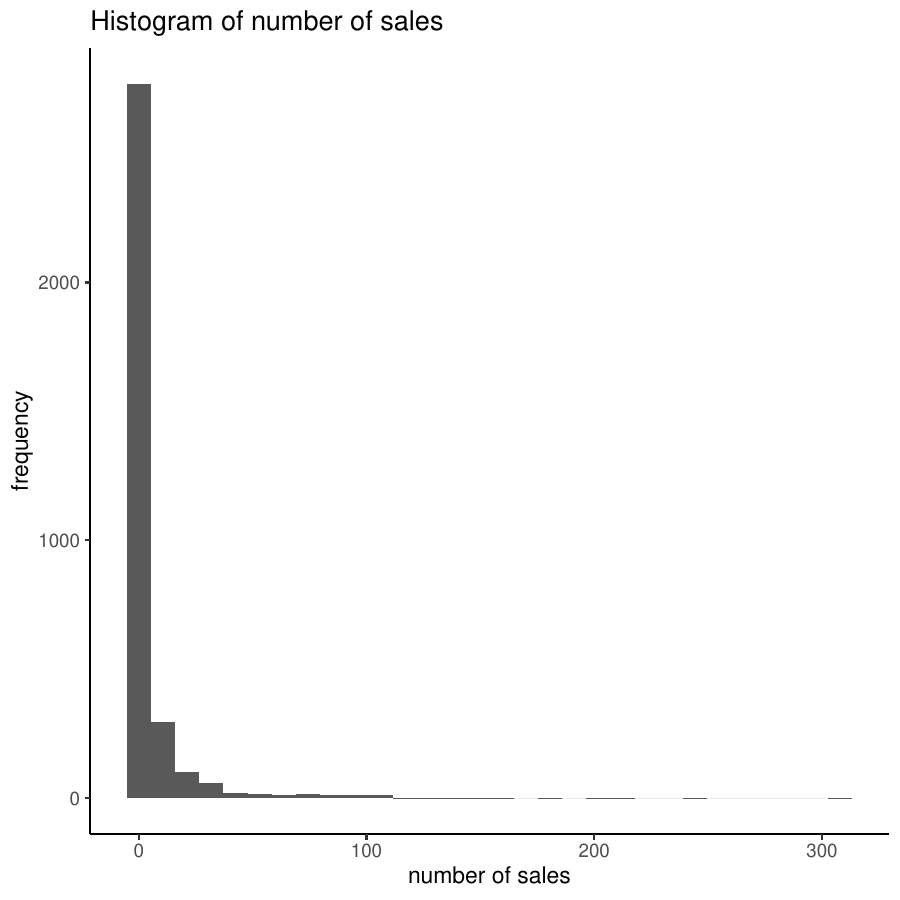}
  \includegraphics[scale = 0.35]{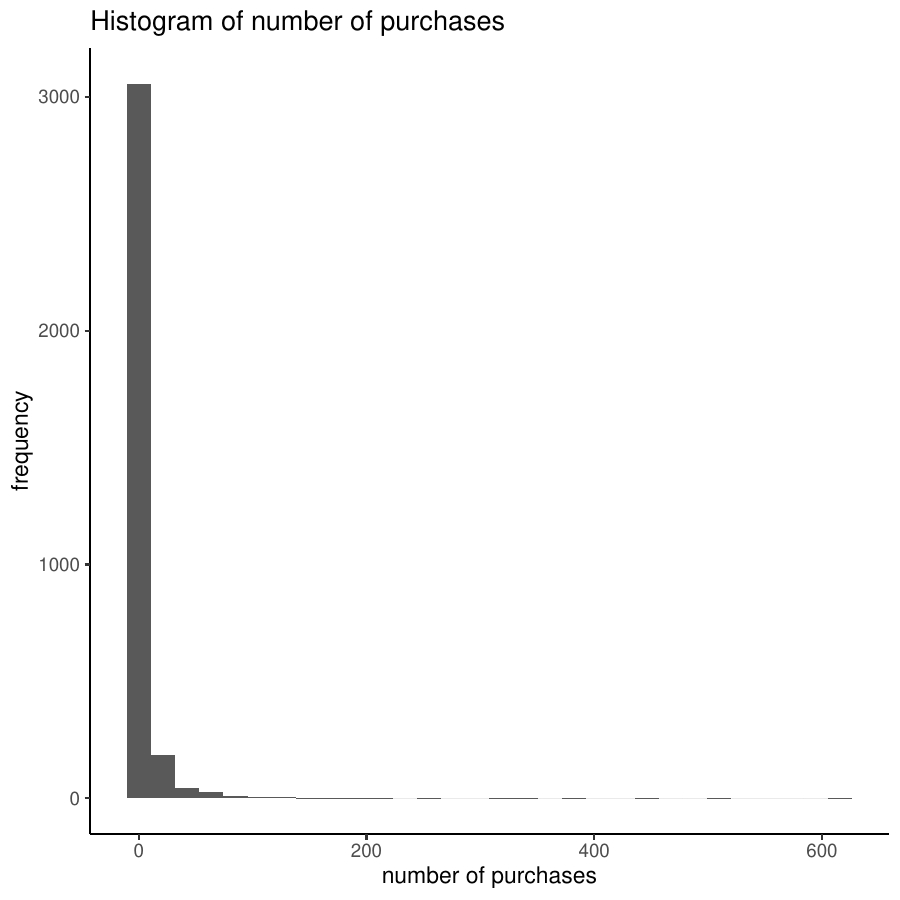}
  \end{center}
	\caption{Histograms of number of sales and purchases for users on SuperRare showing a long-tail distribution. Most users bought and sold few pieces but there are some top collectors and artists that made a lot of purchases and sales.}
	\label{fig:hist-sell-buy}
\end{figure}

SuperRare is a peer-to-peer marketplace for non-fungible tokens (ERC-721 NFTs) built on Ethereum blockchain. More plainly, SuperRare is a marketplace to collect and trade unique, single-edition digital artworks. Each artwork is authentically created by an artist in the network, and tokenized as a crypto-collectible digital item that you can own and trade. SuperRare is one of the earliest crypto art galleries (it started in April, 2018) and is among the most important crypto art marketplaces, by popularity and volume of exchanged artworks. As of today (13 October, 2021), these are some figures for the gallery: 

\begin{itemize}
\item number of tokenized artworks: 29,369;
\item number of sold artworks 17,873 (61\%);
\item sale volume: 66,746.82 ETH or 121,364,200 USD (change rate at transaction time);
\item number of active users: 4412;
\item number of users that created at least one artwork: 1093 (25\%);
\item number of users that sold at least one artwork: 2104 (48\%);
\item number of users that bought at least one artwork: 3372 (76\%).
\end{itemize}

The SuperRare dataset was acquired from the gallery's API and is available on Kaggle.\footnote{\url{https://www.kaggle.com/franceschet/superrare}}. All analyses were conducted in R, taking advantage of the \texttt{tidyverse} packages \cite{W16}.

On SuperRare we can identify four main roles for active users (see Figure \ref{fig:sell-buy}): 

\begin{enumerate}
\item \textit{by-standers}: these are lazy users that buy and sell few artworks (bottom-left part of Figure \ref{fig:sell-buy});
\item \textit{pure sellers}: these are users that sell frequently and buy rarely, they correspond to established artists that create art to sell it (bottom-right part of Figure \ref{fig:sell-buy});
\item \textit{pure buyers}: these are users that buy frequently and sell rarely, they correspond to collectors that buy art to collect and enjoy it or patrons that acquire art to support the artists (top-left part of Figure \ref{fig:sell-buy});
\item \textit{traders}: these are users that buy and sell frequently, they probably buy to re-sell art to make a profit
(top-right part of Figure \ref{fig:sell-buy}).
\end{enumerate}

With respect to the HITS method proposed in this paper for rating users in crypto art, we expect that pure sellers (artists) correspond to nodes with high authority score and pure buyers (collectors) map to nodes with high hub score. On the other hand, traders will have high values for both authority and hub scores. In fact, the product of authority and hub scores is a measure to spot the most relevant art flippers. 

Notice from Figure \ref{fig:sell-buy} that the density of points decades moving from 0 to larger values both vertically and horizontally along the buy and sell dimensions. This hints a relatively high concentration of number of sales and purchases among few sellers and buyers. The intuition is confirmed in Figure \ref{fig:hist-sell-buy}, where we can appreciate the typical long-tail distribution for both sell and buy dimensions. Many users, the so-called trivial many, make few sales or purchases, but a 
significant part of users, the so-called vital few, sell and buy a lot.

\begin{figure}
	\begin{center}
  \includegraphics[scale = 0.28]{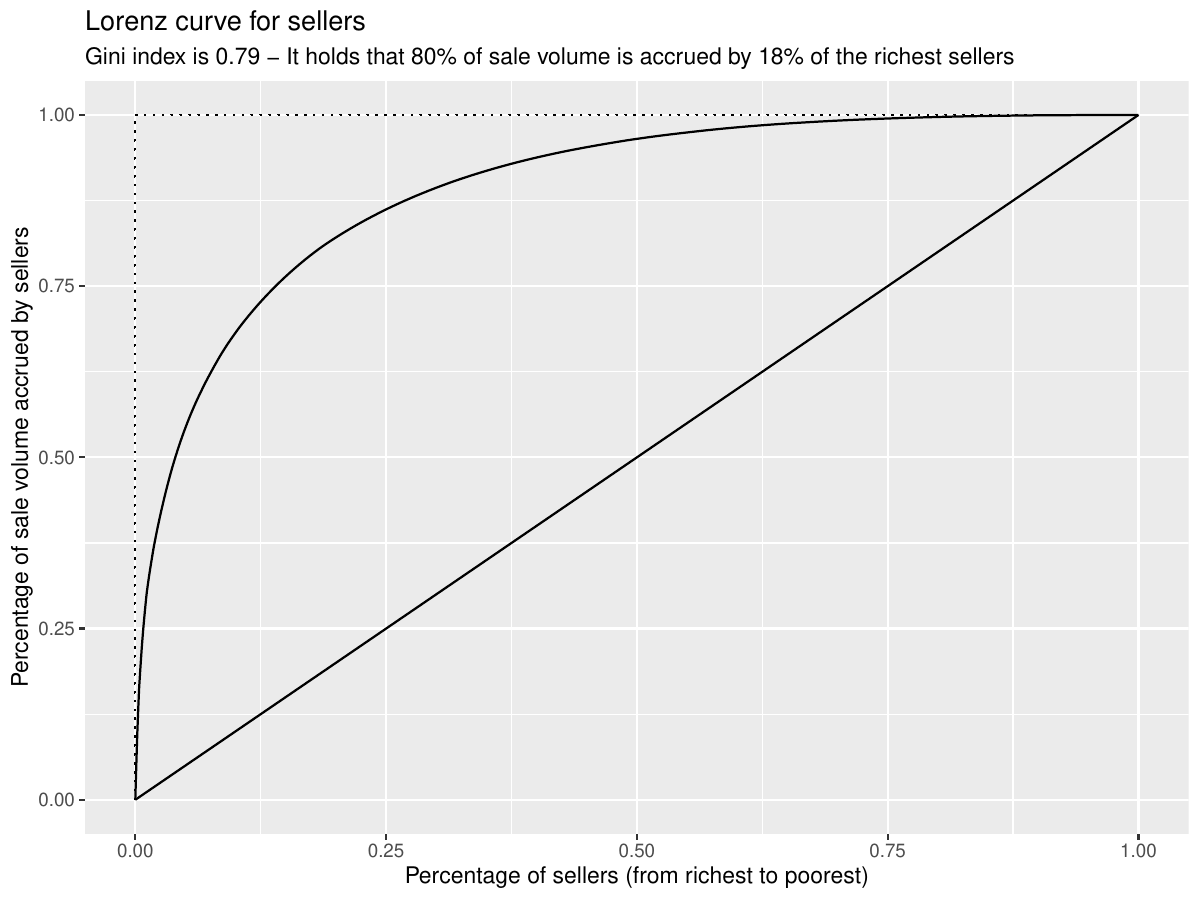}
  \includegraphics[scale = 0.28]{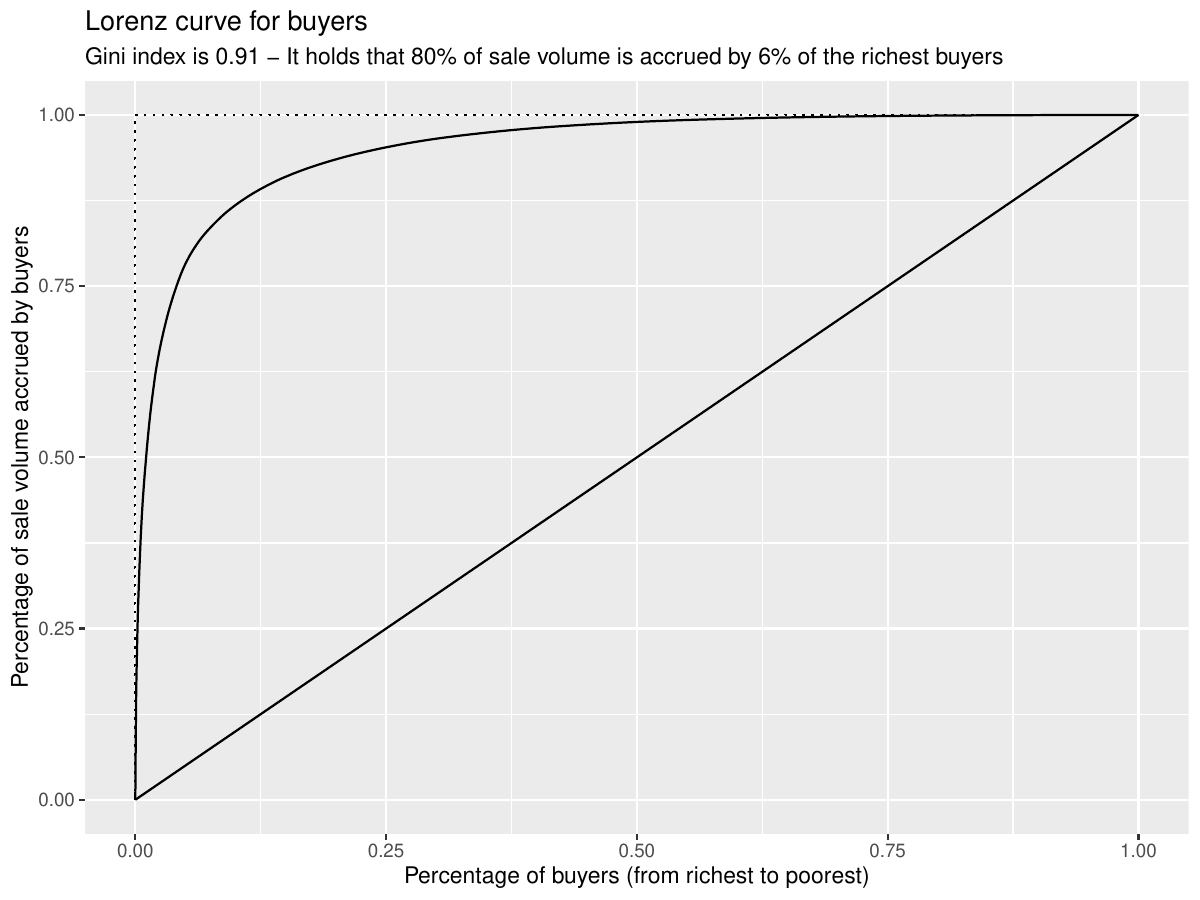}
  \end{center}
	\caption{The Lorenz curves (and Gini indexes) show how the sale volume is concentrated in the hands of top artists and top 
	collectors. }
	\label{fig:lorenz-sell-buy}
\end{figure}

Is this concentration confirmed when we move from number (of sales or purchases) to volume in USD received by artists or spent by collectors? To investigate volume concentration we apply the econometrics tools of Lorenz curve and Gini index (see Figure \ref{fig:lorenz-sell-buy}). When applied to measure the concentration of income within a population:

\begin{itemize}
\item a Lorenz curve plots the cumulative percentages of total income received against the cumulative percentage of number of 
recipients, starting with the richest individual; 
\item the Gini index measures the area between the Lorenz curve and a hypothetical line of absolute equality, expressed as a percentage of the maximum area under the line. Thus a Gini index of 0 represents perfect equality, while an index of 1 implies perfect inequality. 
\end{itemize}

Here, we apply the Lorenz curve and the Gini index to measure the concentration of the sale volume -- the total amount of 
USDs payed to buy art on the gallery -- among few top artists and collectors. It turns out, quite disappointingly, that the crypto art market is highly concentrated among few sellers and even rarer buyers:

\begin{itemize}
\item as for sellers, 80\% of the sale volume is dominated by 18\% of the richest sellers, with a Gini index of 0.79;
\item as for buyers, 80\% of the sale volume is dominated by 6\% of the richest buyers, with a Gini index of 0.91;
\end{itemize}

As a comparison, in the United States, the Gini index\footnote{\url{https://www.prb.org/usdata/indicator/gini/snapshot}} of household incomes was 0.48 in 2014-2018.

The high concentration among top sellers and buyers implies that in the HITS rankings we will have clear winners. Using the weighted version of the HITS metric, the authorities of the network are the top artists that sold for high prices to top collectors, and the hubs are the top collectors that bought for high prices from top artists. These prominent actors are neatly separated from the rest in the HITS rankings, creating a crypto caste that might influence the market.\footnote{This somewhat contrasts with the original spirit of crypto art -- the opposition to a traditional art system highly dominated by a few individuals -- as well as with the original ethos of blockchain -- the ideas of decentralization and democratization.}

Since the role of artists and collectors emerge clearly in the analysis of the SuperRare marketplace, in the following we focus deeper on them. For each role --  artist and collector -- we have identified four network centrality metrics: in-degree, 
in-strength, unweighted authority, and weighted authority for artists and out-degree, out-strength, unweighted hub, and weighted hub for collectors. 

\begin{figure}[t]
	\begin{center}
  \includegraphics[scale = 0.5]{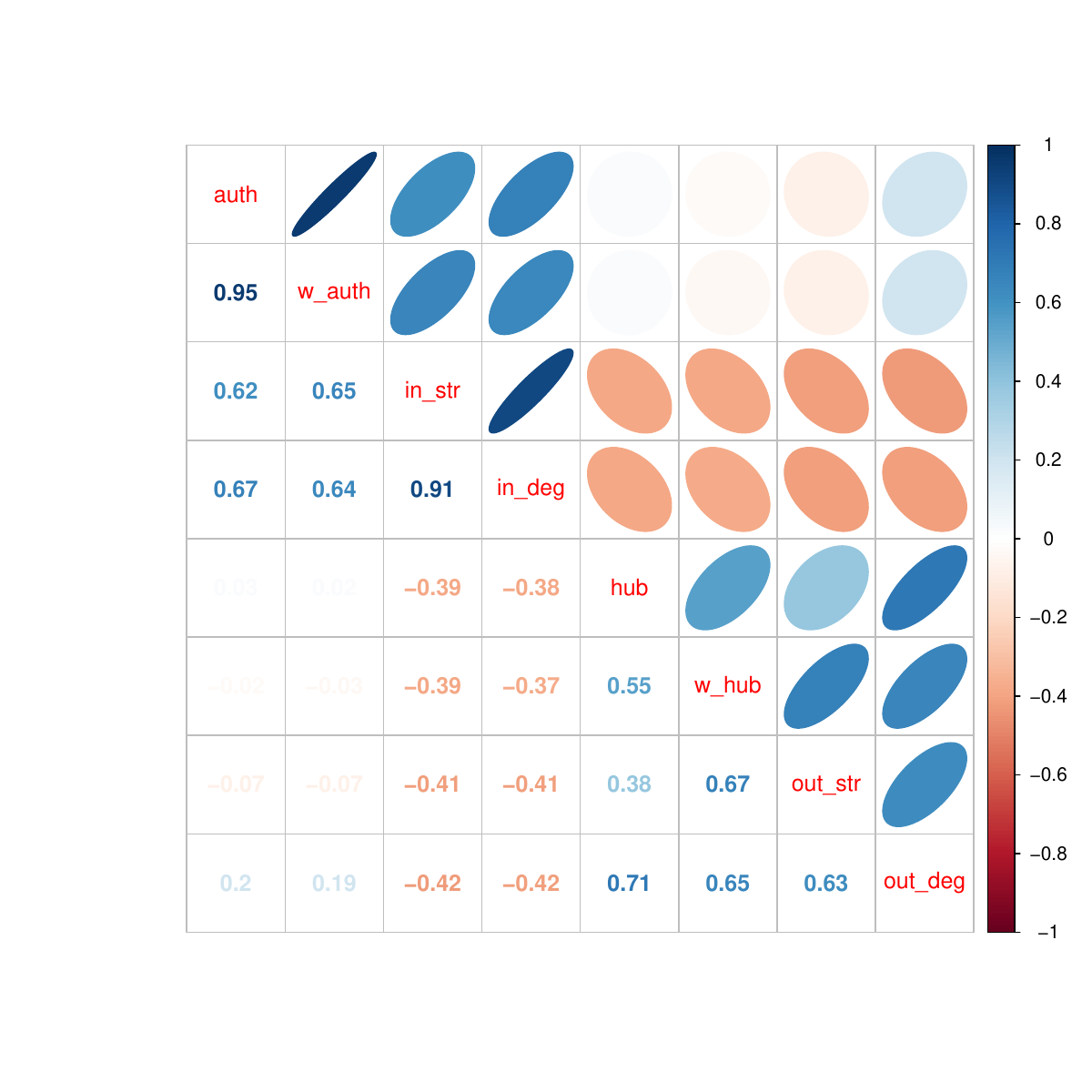}
  \end{center}
	\caption{The Kendall correlation plot among network centrality measures authority (auth),  weighted authority (w-auth), 
	in-strength (in-str), in-degree (in-deg), hub, weighted hub (w-hub), out-strength (out-str), and out-degree (out-deg). }
	\label{fig:corplot}
\end{figure}

First of all, Figure \ref{fig:corplot} contains the correlation plot/matrix for all the eight measures (we used Kendall correlation since the data are highly right skewed). Notice that there are two main clusters of metrics: the artist cluster (in-degree, 
in-strength, unweighted authority, and weighted authority) and the collector cluster (out-degree, out-strength, unweighted hub, and weighted hub). Furthermore, the authority/hub metrics we propose in this paper are not strongly correlated to degree and strength, hence they are telling us something different. Indeed, an artist can sell many pieces, even for a good price, but to unknown collectors, hence collecting a mediocre authority score. On the other hand, an artist can sell a handful of artworks but to the top collectors, therefore easily climbing the ranks of authority. Similarly for collectors.

We can combine the four metrics for artists and collectors in order to better categorize each artist and collector. Let us consider the role of artist, for instance. Suppose we divide the ranking for each measure into three segments: C (low level), for actors with scores in percentile interval [0. 0.5], B (intermediate level) for actors with scores in percentile interval (0.5, 0.9], and A (high level) for actors with scores in percentile interval (0.9, 1]. Hence, each artist is identified by a quadruple of levels, one level for each metric in-degree, in-strength, unweighted authority, and weighted authority, for instance AABC or BBCC. We have in total $3^4 = 81$ possibilities, some of them are very informative on the type of actor at hand. Here are some interesting combinations:

\begin{itemize}
\item AC**, that is an artist with high in-degree and low in-strength. The artist sells a lot but at low prices, collecting a low sale volume. 

\item CA**, that is an artist with low in-degree and high in-strength. The artist sells few pieces but at high prices, gathering a high sale volume. 

\item A*CC, that is an artist with high in-degree and low (unweighted) authority. The artist sells frequently but to minor 
collectors.

\item C*AA, that is an artist with low in-degree and high (unweighted) authority. The artist sells scarcely but to major 
collectors.

\end{itemize}

We can have a similar categorization for collectors, for examples collectors that buy few expensive artworks from important artists and those that buy a lot of cheap artworks from unimportant artists. 

These categories might be useful to devise investment strategies for collectors or marketing strategies for artists. For instance, a low-risk/low-return investment strategy for a collector might prefer \textit{blue-chip artists} that typically sell expensive pieces to top collectors (the AAAA type). On the other hand, a high-risk/high-return investment strategy for a collector might focus on \textit{rising star artists}, that are artists that sold few, inexpensive pieces but to prominent collectors (the CCAB type). There is a chance that these artists have just been discovered from a few far-sighted collectors but are still not popular, hence they represent an investment with a potential large return. On the other hand, an artist might want to target a given category of collectors for marketing purposes. For instance collectors that only buy few expensive artworks from central artists (CABA type), or those that buy many cheap artworks from peripheral artists (ACCC type). 

We can even join the artist metrics with the collector metrics (8 measures in total) to have more insights on the type of user at hand. For instance a user that has high scores on both authority and hub metrics is a top trader, that is, a prominent investor that buys from top artists to resell to top collectors (and make a profit). A user that has high scores for in- and out-degree but low values for authority and hub is a small trader, that is, a modest investor that buys from unknown artists and resells to 
unimportant collectors.

A finer categorization normalizes each metric in interval [0,1] by dividing each score by the maximum score for the metric. Hence an actor, either an artist or a collector, is now defined by a 4-dimensional vector $x = (x_1, x_2, x_3, x_4)$, the $x_i$ are numbers in [0,1]. For instance the vector (0.99, 0.98, 0.95, 0.96) might represent a blue-chip artist while (0.08, 0.04, 0.91, 0.60) might be associated to a rising star artist. Again, we can also consider general users and associate them with a 8-dimensional vector including all the metrics. This allows us to spot different types of traders, as said above. 

\begin{figure}[t]
	\begin{center}
  \includegraphics[scale = 0.5]{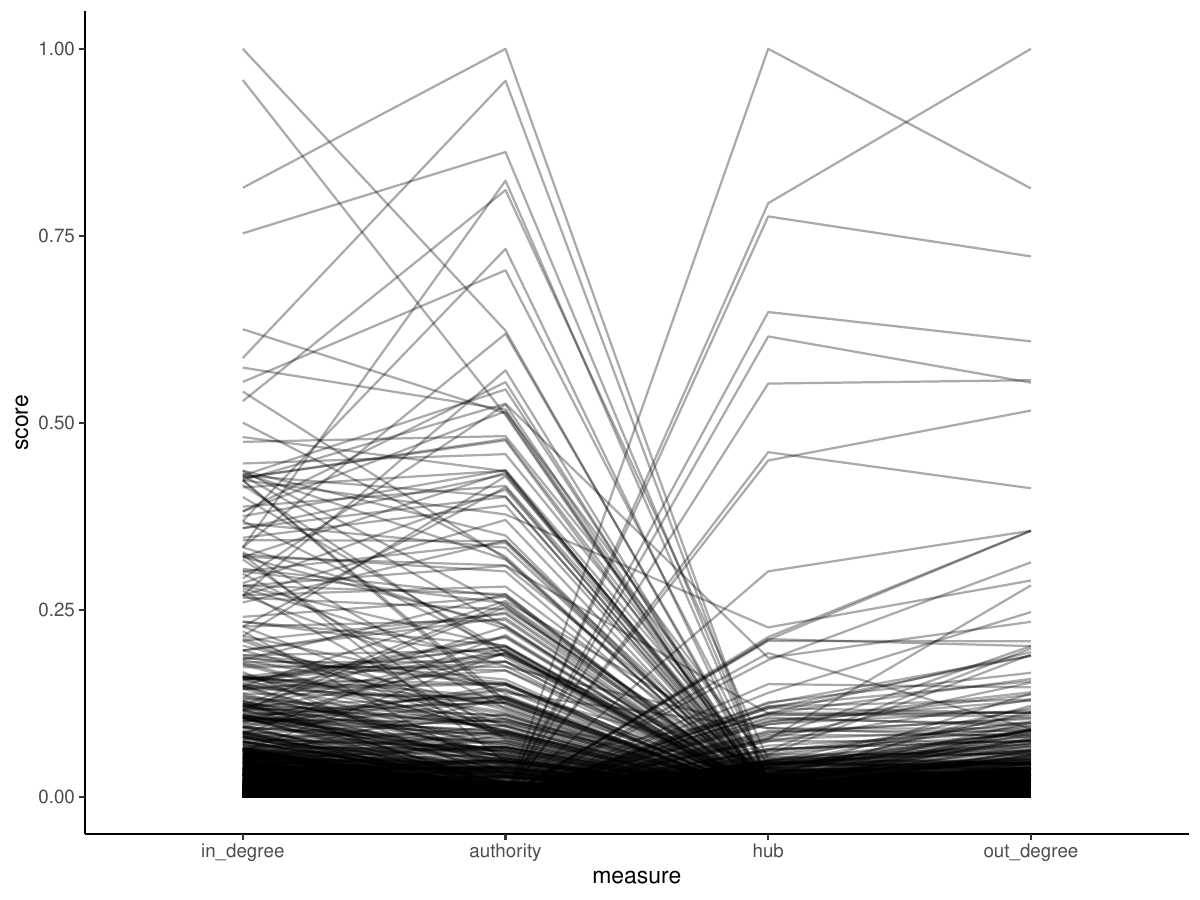}
  \end{center}
	\caption{A one-line-per-user score plot. Each user is represented with four scores for measures in-degree, authority, hub and 
	out-degree and depicted with a line connecting these scores.}
	\label{fig:lined_measures}
\end{figure}

Figure \ref{fig:lined_measures} illustrates this representation for each user in the gallery. For the sake of simplicity, we only show the measures in-degree, authority, hub, and out-degree and connect each score with a line representing the user. Notice the crossing of segments connecting authority and hub in the middle of the figure: typically, users with high authority have low hub and vice versa, since they correspond to different roles (artists and collectors). However, some exceptions exist, and they correspond to traders, having a moderate value for both measures. Also, there are quite a few users that swap positions in the in-degree and authority rankings, for instance artists that sell a lot (high in-degree) but to unknown collectors (low authority) or that sell a few (low in-degree) but to important collectors (high authority). The rankings for collectors (out-degree and hub) are instead more stable. 

\section{Conclusion} \label{sec:conclusion}

We found that the blockchain art market, here represented by the major crypto art gallery SuperRare, is clustered around artists and collectors, although there exist some notable traders. Moreover, we noticed that sales are strongly concentrated among few prominent figures. We hence propose to adapt the Kleinberg's authority/hub HITS method to rate artists and collectors in crypto art. 

Evaluated on the collector-artist network of SuperRare gallery, we found that the proposed method is weakly associated with other network science centralities (degree and strength). We hence propose to cluster active users of the gallery using their scores for these measures and suggest to use this clustering to develop investment strategies for collectors and marketing strategies for artists. 

\bibliographystyle{plain}

\end{document}